\documentclass{article}
\usepackage[utf8]{inputenc}

\usepackage{amsmath}
\usepackage{graphicx}
\usepackage{subfigure}
\usepackage{array}
\usepackage{multirow}
\usepackage{amssymb}
\usepackage{authblk}
\usepackage{xcolor}

\newcommand{\R}{\ensuremath{\mathbb{R}}}
\newcommand{\E}{\ensuremath{\mathbb{E}}}
\newcommand{\norm}[1]{\left\Vert #1\right\Vert}

\DeclareMathOperator*{\rank}{rank}

\DeclareMathOperator*{\F}{F}
\DeclareMathOperator*{\Var}{Var}

\DeclareMathOperator{\G}{\mathcal{G}}

\DeclareMathOperator*{\SNR}{SNR}

\DeclareMathOperator{\Err}{Err}

\definecolor{amber}{rgb}{1.0, 0.75, 0.0}
\definecolor{amethyst}{rgb}{0.6, 0.4, 0.8}
\definecolor{azure}{rgb}{0.0, 0.5, 1.0}
\definecolor{aqua}{rgb}{0.0, 1.0, 1.0}
\definecolor{asparagus}{rgb}{0.53, 0.66, 0.42}
\definecolor{applegreen}{rgb}{0.55, 0.71, 0.0}

\newtheorem{theorem}{Theorem}[section]
\newtheorem{lemma}[theorem]{Lemma}
\newtheorem{remark}[theorem]{Remark}
\newtheorem{definition}[theorem]{Definition}
\newtheorem{example}[theorem]{Example}
\newtheorem{corollary}[theorem]{Corollary}
\newtheorem{proposition}[theorem]{Proposition}

\title{On the reconstruction accuracy of multi-coil MRI with orthogonal projections}
\author[1]{Anna Breger}
\author[2]{Gabriel Ramos Llorden}
\author[3]{Gonzalo Vegas Sanchez - Ferrero}
\author[3]{W. Scott Hoge}
\author[1]{Martin Ehler}
\author[4]{Carl-Fredrik Westin}
\date{September 2019}
\affil[1]{\footnotesize Department of Mathematics, University of Vienna, Austria}
\affil[2]{\footnotesize Department of Psychiatry, Brigham and Women's Hospital, Harvard Medical School, Boston, MA}
\affil[3]{\footnotesize Department of Radiology, Brigham and Women's Hospital, Boston, MA}
\affil[4]{\footnotesize Laboratory of Mathematics in Imaging, Brigham and Women's Hospital, Harvard Medical School, Boston, MA}

\begin{document}

\maketitle

\begin{abstract}
MRI signal acquisition with multiple coils in a phased array is nowadays commonplace. The use of multiple receiver coils increases the signal-to-noise ratio (SNR) and enables accelerated parallel imaging methods. Some of these methods, like GRAPPA or SPIRiT, yield individual coil images in the k-space domain which need to be combined to form a final image. Coil combination is often the last step of the image reconstruction, where the root sum of squares (rSOS) is frequently used. This straightforward method works well for coil images with high SNR, but can yield problems in images with artifacts or low SNR in all individual coils.  We aim to analyze the final coil combination step in the framework of linear compression, including principal component analysis (PCA). With two data sets, a simulated and an in-vivo, we use random projections as a representation of the whole space of orthogonal projections. This allows us to study the impact of linear compression in the image space with diverse measures of reconstruction accuracy. In particular, the $L_2$ error, variance, SNR, and visual results serve as performance measures to describe the final image quality. We study their relationships and observe that the $L_2$ error and variance strongly correlate, but as expected minimal $L_2$ error does not necessarily correspond to the best visual results. In terms of visual evaluation and SNR, the compression with PCA outperforms all other methods, including rSOS on the uncompressed image space data.
\end{abstract}

\section{Introduction}
Magnetic Resonance Imaging (MRI) is a unique medical imaging modality that provides excellent image quality without ionizing radiation, but on the other hand is relatively slow. The acquired data, % samples,
sampled in the k-space domain, % also known as the k-space data, are samples of
corresponds to the Fourier transform of the spatial-domain MR image. To reconstruct
an accurate MR image,  % indicates
the total amount of k-space data that must be
acquired to avoid reconstruction artifacts is dictated by sampling theory.
% WSH: reducing the data-load is an active area of MRI research.  I re-wrote the above to reflect the theory more generally.
%      And, the next sentence is difficult to reconcile with the history of parallel imaging.
% As this number is relatively large, and cannot be arbitrarily reduced,  
% the total scan time cannot be shortened there are without compromising the image quality.

A turning point in MRI reconstruction was the implementation of phased-array coils in the acquisition pipeline. With phased-array coils, the k-space data is acquired in multiple receivers/coils simultaneously. On top of increasing the signal-to-noise (SNR) ratio, a phased-array coil acquisition allows accelerating the total acquisition time \cite{doi:10.1002/mrm.1910160203}. These methods, known as Parallel Imaging (PI) methods, exploit the fact that there exist complementary k-space data information in each coil/receiver. In PI, the k-space data set is undersampled, but the multiple measurements allow % the reconstruction of
missing k-space samples to be recovered through an % in a kind of
inverse problem reconstruction framework. See \cite{overview} for an overview of basic reconstruction algorithms for PI and their history. 

The computational costs and required memory of the reconstruction algorithms highly depend on the dimension of the phased array, i.e. the number of receiver coils. Several coil compression methods have been developed that reduce to a smaller set of virtual channels without a significant loss of SNR, see e.g.~\cite{doi:10.1002/mrm.24267},\cite{10.1002/mrm.26032},\cite{feas}. Moreover, PCA-based methods have shown to even have a beneficial denoising effect, e.g. \cite{kpca}. In \cite{doi:10.1002/mrm.21237}, the optimal linear projection for coil compression based on the resulting SNR is derived. Note that this is related to our analysis approach, but we work in the image domain % space
instead of the k-space domain.

After the obligatory compression step, PI reconstruction methods can be applied and generate results in different output spaces. Some of the methods operate in % reconstruct the image directly
the image domain, % (image-based methods),
e.g. SENSE \cite{sense}.  Other methods operate in the k-space domain % based methods retrieve the missing k-space data, thereby
and generate % providing
a fully-sampled k-space array per each receiver/coil, e.g. GRAPPA or SPIRiT \cite{doi:10.1002/mrm.10171,RIS_0}.
% Next, all k-space data is inversely Fourier transformed and the resulting image array has to be  combined into a single final image.
To form a spatial-domain image that can be
% This image represents the conventional MR image and is
used for analysis, quantification, or visualization purposes, the k-space domain data is 
inversely Fourier transformed and the resulting % image array
individual coil images % has
need
to be combined into a single final image.

%Importantly, there exist multiple techniques to combine the reconstructed images in each coil. These techniques may impact on the quality of subsequently analysis/estimation step differently. In fact, there is a lack of consensus about the best approach to combine such multiple information.

In \cite{doi:10.1002/mrm.1910160203}, optimal methods to combine the arrays from phased array elements have been developed. These methods rely on detailed knowledge of the receive sensitivity of each coil. % 's magnetic fields.
In practice, the exact position and sensitivity of each coil is not always known or not possible to
estimate because of physical limitations. % computational
To overcome this issue, a method that combines the data without detailed knowledge of the coils,
while preserving a high SNR, is desirable. % needed.
% For this purpose,
Because of this, the root sum of squares (rSOS) method has become the standard method of combining multi-coil images in MRI \cite{doi:10.1002/mrm.1910160203, doi:10.1002/(SICI)}. For arrays
with high SNR, the rSOS yields nearly optimal reconstruction, whereas problems arise if all coils yield low SNR. Especially data with artifacts are problematic since rSOS weights them equally to the
non-defective parts.

Noise is ever present % has a % huge impact
in MRI imaging and is dominated by two sources: % often caused by
thermal noise in the receiver apparatus and the physiological noise from patient's body itself. %  due to radiofrequency emissions. 
Moreover, the SNR in the coil images depends highly on the field strength, scanner hardware, and data acquisition % imaging
modality (e.g. T1/T2/diffusion weighted). Extensive statistical noise analysis can be found in \cite{noise}. In \cite{den1,den2}, PCA was used to jointly reconstruct multi-coil data from diffusion MRI exploiting the expected redundancy of data acquired with different gradient directions. Under the framework of random matrix theory, authors derived practical rules to accurately nullify noise-only principal components. This way, dMRI data can be effectively denoised while preserving the important information.

In this work, we aim to analyze the coil combination in a comprehensive way by comparing different performance measures. To address coil combination independent of prior knowledge, we will study the rSOS in the framework of linear compression via orthogonal projections. Random projections will serve us as a tool to study the correlation between $L_2$ reconstruction error and voxel variance of the varying reconstructions. Correlation analysis with random orthogonal projections has been studied in a related context in \cite{ortho}, yielding an underlying understanding of the relation between important information preservation features in data combination and compression. 

Besides the correlation analysis, we observe that optimal $L_2$ reconstruction error, i.e. mean squared error (MSE), does not necessarily correspond to optimal visual evaluation of the reconstructed images. This corresponds to described problems when measuring image quality by the MSE, see e.g. \cite{imqual, qualmse, ssim}. Nevertheless, the often used peak signal-to-noise ratio (PSNR) is based on the computation of the MSE. Here, we will additionally evaluate the reconstructions regarding the SNR and visualization of the images. The results confirm an improvement by compressing image space data with PCA.     

The outline is as follows: first we will explain how rSOS can be interpreted in the context of orthogonal projections. Moreover, we will describe samples of random orthogonal projections, that shall serve us as coverings enabling numerical experiments. Then we describe the two data sets, a simulated and an in-vivo, that we use for experimental investigations. The simulated data set allows us to compare the behavior with different noise levels in the coils. In the results section, we show scatter plots describing the relation between the $L_2$ reconstruction error and the voxel variance in reconstructed magnitude images, as well as comparing it with the SNR and visual outcomes. Finally, we will interpret and discuss the results in Section \ref{discussion}.

\section{Methods}
We aim to study the impact of linear compression in coil combination with rSOS. Such a combination step is necessary for phased-array data that has been processed with GRAPPA, SPIRiT, or other PI methods in k-space. Before combining the fully sampled, multidimensional image space data as the final step of the reconstruction pipeline, we include linear compression with orthogonal projections. 

The common reconstruction pipeline including our linear compression can be summarized as follows:
\begin{equation*}
\hat{y}_i \in \mathbb{C}^d \ \xrightarrow{\text{\tiny{GRAPPA}}} \ \hat{x_i} \in \mathbb{C}^d \ \xrightarrow{\text{\tiny{IFFT + abs}}} \ x_i \in \mathbb{R}^d \xrightarrow{\text{\tiny{projection}}} \ p x_i \in \mathbb{R}^k \ \xrightarrow{\text{\tiny{rSOS}}} \ \norm{px_i}_2 \in \R
\end{equation*}
where $\hat{y}_i$ corresponds to an undersampled k-space voxel, $\hat{x}_i$ is fully sampled in k-space after some PI reconstruction (e.g. GRAPPA) and $x_i$ is the image space data. Our analysis takes place on the image space data in $\mathbb{R}^d$ that is subsequently projected to $\mathbb{R}^k$ with $k<d$.

\subsection{Framework of orthogonal projections}
As the basis of our analysis we will use random orthogonal projections yielding different linear coil compressions. We will work with image data consisting of $d$ channels that shall be combined into one final magnitude image. To do so, we study the space of $k$-dim linear subspaces of $\R^d$, which can be identified by orthogonal projections
\begin{equation}
    \G_{k,d} = \{p \in \R^{d \times d}: p^2 = p, p^T = p, \rank(p) = k \},
\end{equation}
called the Grassmannian. 

Let $x = \{x_i\}_{i = 1}^{m} \in \R^d$ be an image space data set with $m$ voxels measured by $d$ coils. Then, the final image volume is given by $\norm{px}_2$ with $p$ in $\G_{k,d}$, where $d$ is the fixed number of  channels and $k < d$ varies. This corresponds to projecting the $d$ coil channels to different dimensions $k$ and computing root sum of squares (rSOS) subsequently. Note that we can study the rSOS itself in this context, since rSOS$(x) = \norm{x}_2$ and it holds for all $p \in \G_{k,d}$ that the expectation value 
\begin{equation}\label{expsos}
    \E[ \norm{px_i}_2] = c \cdot \norm{x_i}_2 \quad \forall x_i \in \R^d,
\end{equation}
with $c = \big ( \Gamma(\tfrac{k+1}{2})  \Gamma(\tfrac{d}{2}) \big )/ \big( \Gamma(\tfrac{k}{2}) \Gamma(\tfrac{d+1}{2}) \big )$, where $\Gamma$ denotes the Gamma function. 

\smallskip
This is based on the Chi distribution and the fact that the length of a random unit vector projected onto a fixed $k$-dimensional subspace has the same distribution as the length of a unit vector in $\R^d$ being projected onto a random $k$-dimensional subspace (see e.g. \cite{doi:10.1002/rsa.10073}).

\begin{remark}
The equality \eqref{expsos} allows us to analyze the rSOS itself in the framework of orthogonal projections, i.e. no added coil compression in the image space. Summing up the projected voxels obtained by a reasonably big sample set of random orthogonal projections $p = \{p_l\}_{l = 1}^{n} \in \G_{k,d}$, yields approximately the rSOS combined voxel up to the constant $c$, i.e.
\begin{equation}
    \tfrac{1}{n} \sum_{l=1}^n \norm{p_l x_i}_2 \approx c \cdot \norm{x_i}_2 .
\end{equation}
Since we linearly rescale the final image volumes between [0,1] to enable error estimation with the ground truth, the constant is negligible.  Note that also principal component analysis (PCA) yields an orthogonal projection that lays in the Grassmannian manifold and therefore can be studied in that context. Moreover, we will use random orthogonal projections, i.e. projections $p \in \G_{k,d}$ distributed according to the orthogonally invariant probability measure $\mu_{k,d}$, as samples of orthogonal projections, see e.g. \cite{doi:10.1080/10586458.2016.1226209}, \cite{BREGER20181}. 
\end{remark}

\subsection{Measures of performance}

The $L_2$ error will serve us as a measure of accuracy, when comparing the ground truth data with the final reconstructions, i.e. the combined image space data $x = \{x_i\}_{i = 1}^{m} \in \R^d$. Note that w.l.o.g. $x$ contains here just the voxels in the region of interest (discarding the background) and not the full image volume. For a fixed projection $p \in \G_{k,d}$, the error between some provided ground truth $y = \{y_i\}_{i = 1}^{m}$ and the combined magnitude image voxels $\norm{px}_2 := \{\norm{px_i}_2 \}_{i = 1}^{m}$, is then given by 
\begin{equation}\label{err}
    \Err(y,\norm{px}_2) := \frac{1}{m} \sum_{i=1}^{m} \big(y_i - \norm{px_i}_2 \big)^2.
\end{equation}
We aim to study the relation between the reconstruction error and the variance in the reconstructed magnitude images, which can be interpreted as contrast in noise-free images. The variance of the final magnitude images depending on the projection method can be computed by 
\begin{equation}\label{var}
    \Var\big(\norm{px}_2 \big) = \frac{1}{m(m-1)} \sum_{i<j} \big( \norm{px_i}_2 - \norm{px_j}_2 \big)^2.
\end{equation}
Moreover, we will use the mean signal-to-noise ratio to interpret the performance. Note that in MRI it is common to use the non-squared version, i.e.
\begin{equation}\label{SNR}
    \SNR\big(\norm{px}_2 \big) = \frac{1}{m}\sum_{i=1}^{m} \frac{\norm{px_i}_2}{\sigma},
\end{equation}
where $\sigma$ corresponds to the estimated standard deviation of the noise, see e.g. \cite{noise}. 

\subsection{Random projections as coverings}\label{rancov}
To enable a numerical analysis, we need a finite set of orthogonal projections that represents the overall space well, i.e. covers the Grassmannian $G_{k,d}$ properly. To measure how well a set covers the underlying space, we use the definition of the covering radius. 
\begin{definition}
Let the \emph{covering radius} of a finite set $\{p_1, \dots, p_n \}\subset \G_{k,d}$ be denoted by
\begin{equation}\label{eq:def cov}
\rho(\{p_l\}_{l=1}^n):=\sup_{p\in \G_{k,d}} \min_{1\leq l\leq n} \|p-p_l\|_{\F},
\end{equation}
where $\|\cdot\|_{\F}$ is the Frobenius norm. 
\end{definition}
Note that the smaller the covering radius, the better the finite set of projections $\{p_l\}_{l=1}^n$ represents the entire space $\G_{k,d}$: it yields smaller holes and the points are better spread. 

Let $\mu_{k,d}$ denote the normalized Riemannian measure on $\G_{k,d}$ as before. According to \cite{10.1093/imrn/rnv342}, the expectation of the covering radius $\rho$ of $n$ random points $\{p_j\}_{j=1}^n$, independent identically distributed according to $\mu_{k,d}$, satisfies \footnote{We use the symbol $\sim$ to indicate that the corresponding equalities hold up to a positive constant factor on the respective right-hand side.}
\begin{equation}\label{eq:random asympt}
\mathbb{E}\rho \sim n^{-\frac{1}{k(d-k)}} \log(n)^{\frac{1}{k(d-k)}}.
\end{equation}

Following the definition of asymptotically optimal covering in \cite{BREGER20181}, the expectation yields an optimal  covering radius up to a logarithmic factor $\log(n)^{\tfrac{1}{k(d-k)}}$. To remain flexible in the dimension of $\G_{k,d}$, i.e. the choice of $k$ and $d$, and the number of projections $n$, we will work here with random projections distributed according to $\mu_{k,d}$ rather than constructing optimal covering sequences as in \cite{BREGER20181}. These random orthogonal projections can be efficiently computed by the $QR$ decomposition of a matrix $M = QR$ with independent standard normal distributed entries \cite[Theorem 2.2.2]{Chikuse:2003aa}.     

In the following we will use finitely many samples of random projections for the experimental analysis. 

\section{Data} 
We will run our numerical analysis on two different MR data sets: a simulated T1-weighted data set from brainweb (\cite{Cocosco97brainweb:online}, \cite{816072}, \cite{712135}) and an in-vivo data set from a head coil receiver.  
\subsection{Simulated data set}\label{simdat}
Simulation experiments were conducted to assess the quality in image reconstruction for different types of projections in a controlled, rigorous manner. To do so, first, a ground-truth volume  was created with the popular numerical simulator BrainWeb \cite{Cocosco1997}.  A (magnitude) multi-slice T1-weighted volume was simulated with a Spoiled Fast Low Angle Shot (SFLASH) sequence with the following parameters:  TR/TE = 20 /10 ms, flip angle of 90 degrees and ETL = 1. Matrix size: $181 \times 21 \times 76 $ with isotropic voxel size of 1 mm. 

Next, simulated images with 32 channels, mimicking a 32-channel coil acquisition, were created. First, synthetic coil sensitivity profiles were simulated assuming a smooth Gaussian profile \cite{noise}. Voxel-wise multiplication of those coil sensitivities by the simulated ground-truth image creates the 32 coil-based images. Uncorrelated complex Gaussian noise with zero mean and standard deviation $\sigma$ was added. Finally, to simulate a magnitude-based acquisition, the absolute value of $32$ noisy images were taken. The value of $\sigma$ was chosen differently to recreate experiments with different noise levels.

\subsection{In-vivo data}\label{invivodat}
MR data was acquired in-vivo from a healthy volunteer using a Siemens (Erlangen, Germany) 3T Prisma
equipped with a $32$-channel head coil receiver.  An Echo-Planar diffusion sequence was employed to
acquire 24 slices in a 2mm iso-tropic volume, with slice-thickness 2mm (TR=3.2 sec, TE=85ms,
flip-angle=90, matrix size 128x128, FOV: 256mm by 256mm). A T2-weighted volume was acquired, with no
diffusion weighting (a "b=0" image), followed by 62 volumes acquired using a single repeated
diffusion vector and diffusion setting of b=1000. The measured EPI data was reconstructed using
Dual-Polarity GRAPPA (DPG) to minimize Nyquist ghosts, see \cite{hoge:mrm2015:dpg}. 
% WSH: I switched the DPG reference to one that is more relevant.  The earlier citation was for the
% multi-band version, which was not used for the in-vivo data you show.
In-plane acceleration of the original data was $R=2$. The ground-truth image was formed by motion-correcting
the 62-volume diffusion-weighted series, using the Advanced Normalization Tools (ANTs) library
\cite{10.1007/978-3-319-14678-2_1}, and averaging across time. The first diffusion direction was
studied regarding linear coil combination and compared to the computed the ground-truth. Few
outliers have been removed by using the $99.99$th percentile and setting the remaining $0.01\%$ to
the maximum of the used percentile.

\section{Results}
Both studied image volumes consist of phased-array data with $32$ channels, therefore we work with projections $p$ in $G_{k,d}$ with $d = 32$ and varying $k$. In the following scatter plots we see how the $L_2$ reconstruction error \eqref{err} relates to the voxel variance \eqref{var} and state corresponding mean SNR values in the visualization. 

Each point in the plots represents a reconstruction with linear compression and rSOS, yielding a final magnitude image volume as described in the previous sections. The symbols + correspond to a projection $p \in G_{k,32}$, i.e. $\norm{px}_2$, and the colors to the different dimensions $k$: 
\[{\color{azure} \textbf{+}} \ p \in \G_{1,32}, {\color{orange} \textbf{+}} \ p \in \G_{4,32},{\color{amber}\textbf{+}} \ p \in \G_{12,32}, {\color{amethyst}\textbf{+}} \ p \in \G_{20,32},{\color{applegreen} \textbf{+}} \ p \in \G_{28,32},{\color{aqua} \textbf{+}} \ p \in \G_{31,32} .\]
As described in Section \ref{rancov} the projections $p$ serve as a covering of the underlying space and are chosen randomly according to the orthogonally invariant probability measure $\mu_{k,d}$. For the simulated data set (see Section \ref{simdat}) $n=500$ projections are randomly chosen for every space $\G_{k,32}$ with $k \in \{1,4,12,20,28,31\}$, for the in-vivo data set (see Section \ref{invivodat}) $n=1000$. 

In the scatter plots the symbol $\square$ corresponds to the rSOS, i.e. $\norm{x}_2$, and $\circ$ to the compression with the orthogonal projection provided by PCA. The colors correspond to the different spaces $\G_{k,32}$ as stated above. 

\begin{figure}
 \subfigure[SNR = 10.64]{\includegraphics[width = 0.5\textwidth]{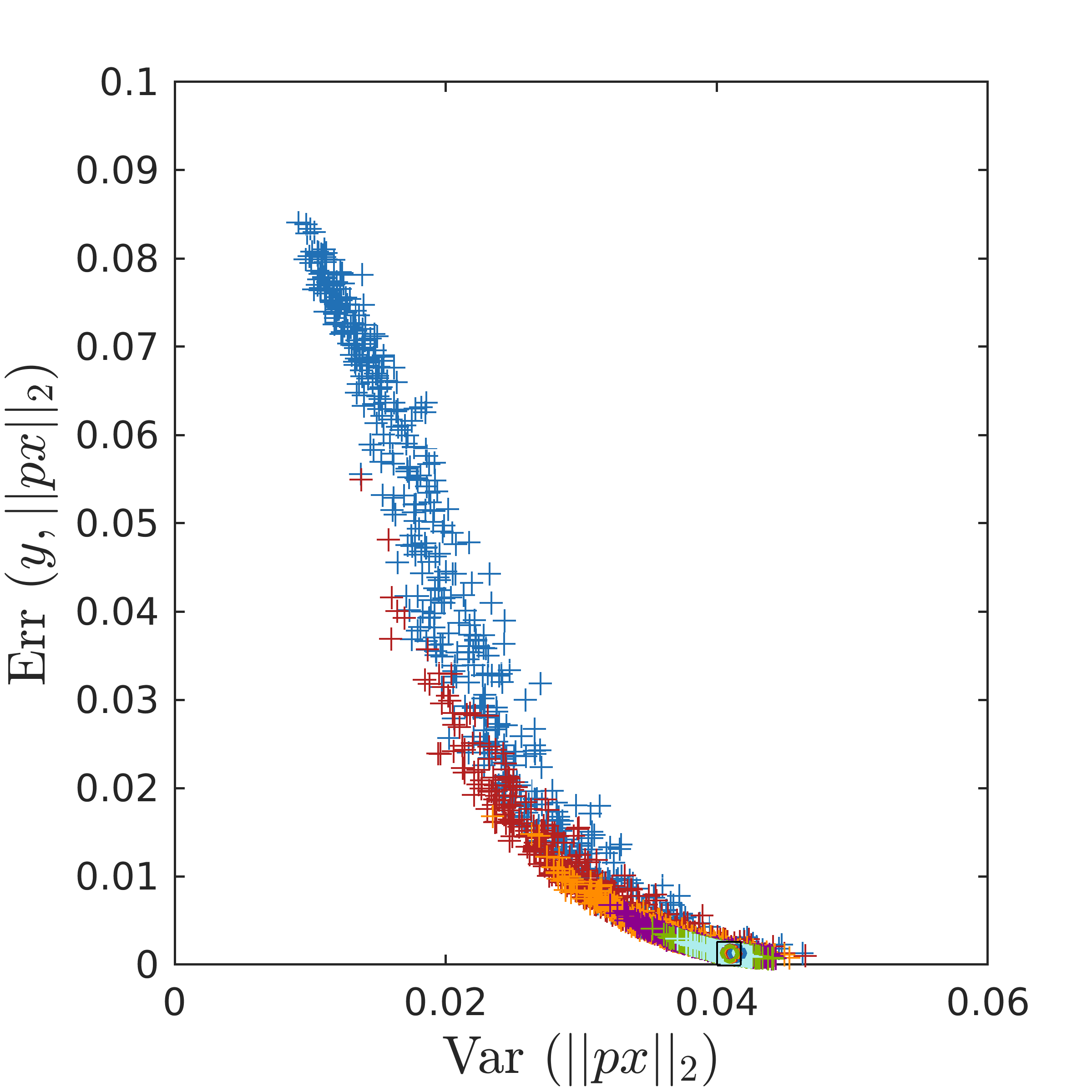}}
  \subfigure[SNR = 5.45]{\includegraphics[width = 0.5\textwidth]{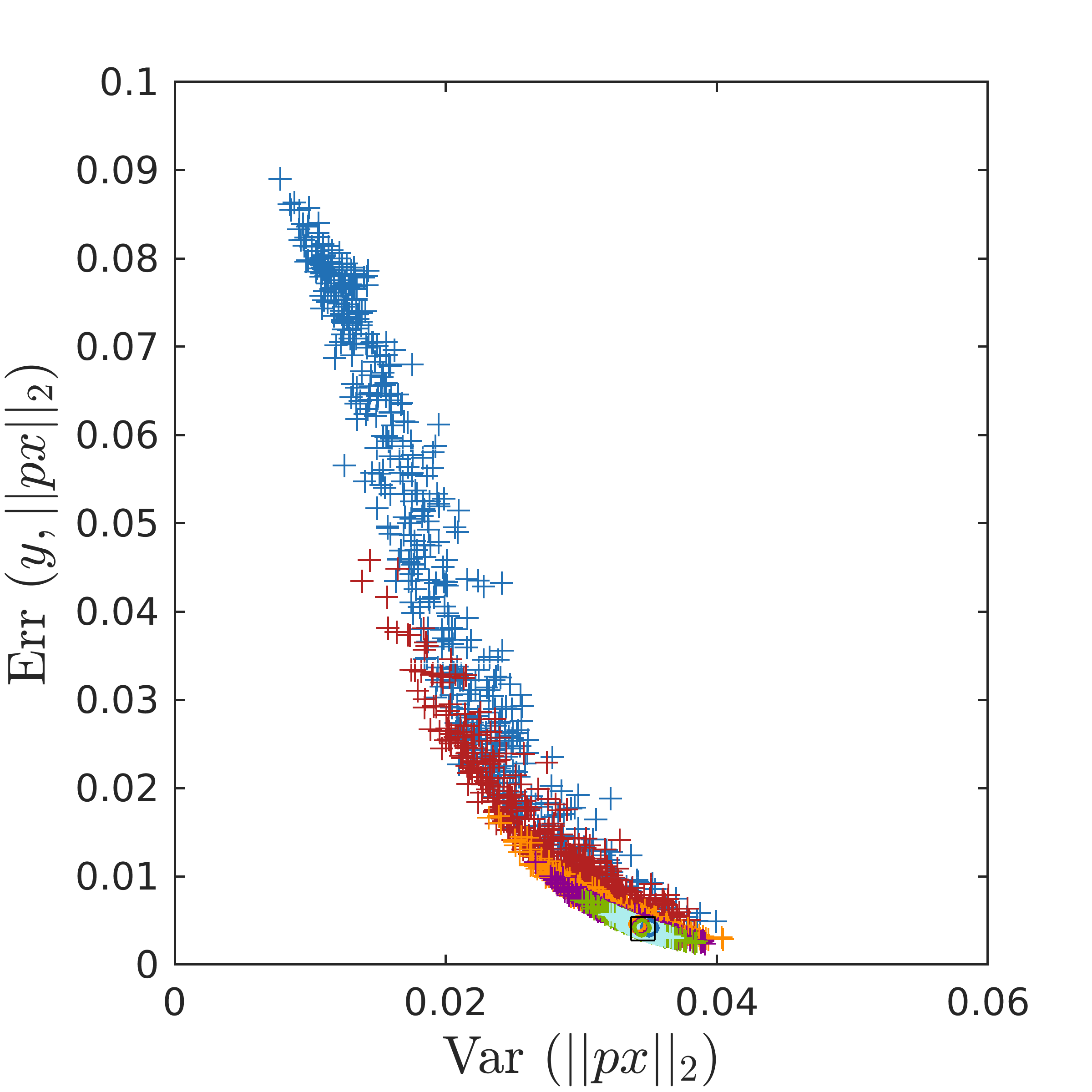}}
 \subfigure[SNR = 2.77]{\includegraphics[width = 0.5\textwidth]{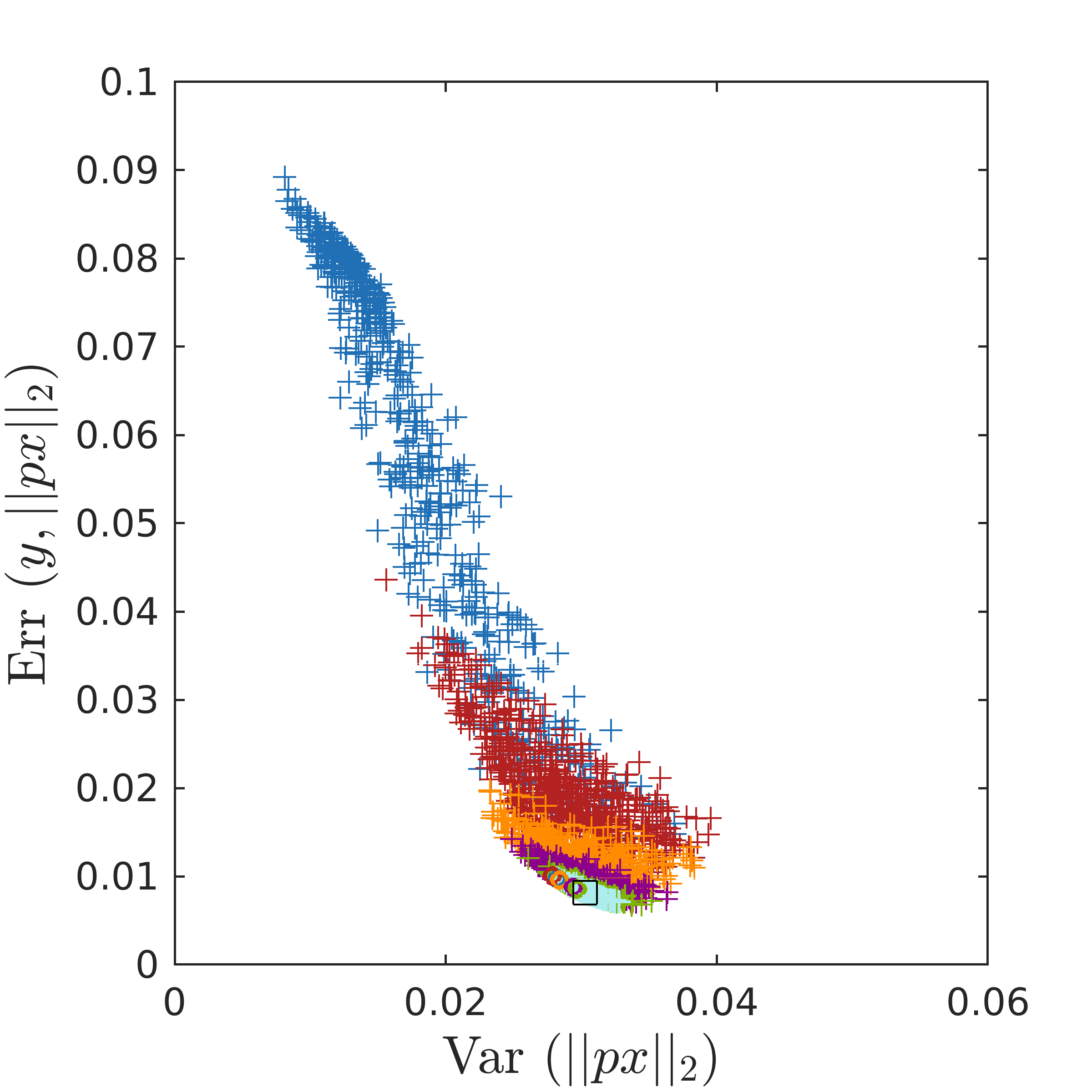}}
 \subfigure[SNR = 1.96]{\includegraphics[width = 0.5\textwidth]{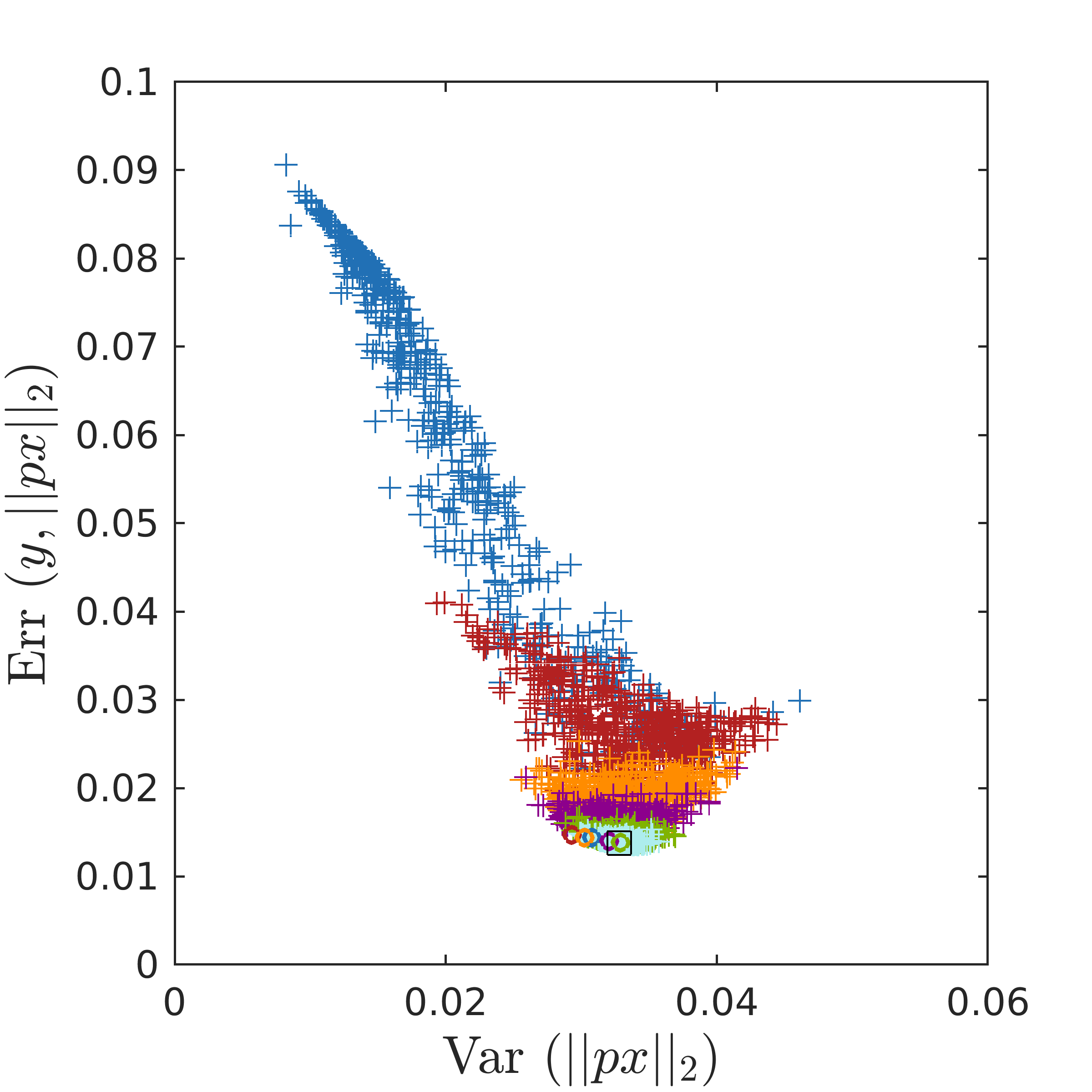}}
    \caption{Scatter plots for the simulated data set with different noise levels, showing the $L_2$ reconstruction error \eqref{err} and variance \eqref{var} of the final images obtained by combining the coils with rSOS and compression by random projections and PCA in $\G_{k,32}$. A varying amount of Gaussian noise was added in each coil assuming no correlation \cite{noise}. The SNR value corresponds to the mean over all voxels in the reconstructed rSOS image volume without the linear compression.}
    \label{simplots}
\end{figure}

\begin{figure}
    \includegraphics[width = 1\textwidth]{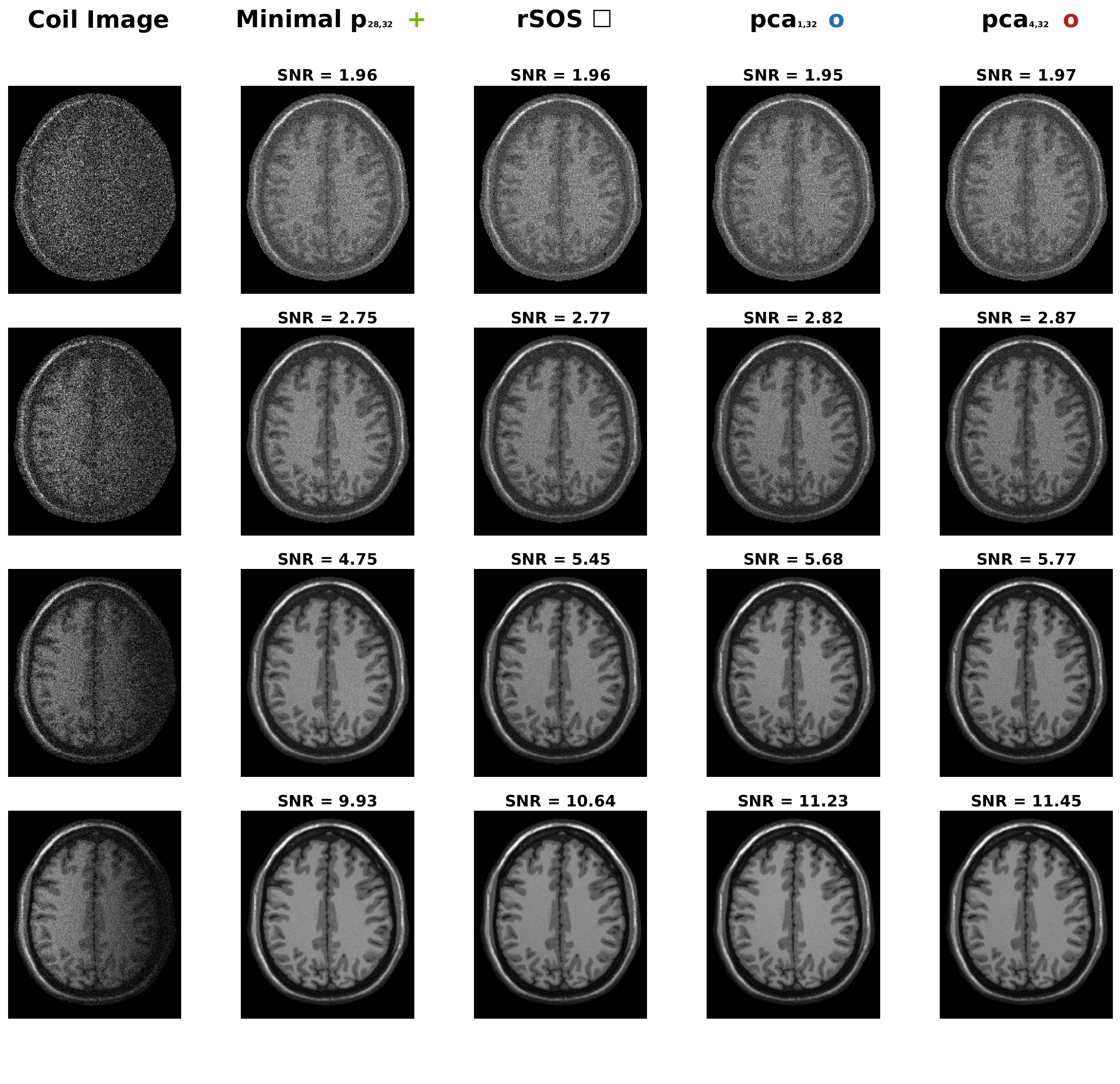}
    \caption{Cross-sectional image corresponding to the scatter plots in Figure \ref{simplots} for the simulated data set with different noise levels. The left column shows the first channel of the simulated noisy $32$-dim coil array before the coil combination step. The second column shows the final image obtained by rSOS including compression with the random projection $p \in \G_{28,32}$ that yields the minimal $L_2$ reconstruction error \eqref{err}. The third column shows the image provided by rSOS without compression. The second last columns show the final images when using PCA in $\G_{1,32}$ and $\G_{4,32}$ for compression, yielding the highest mean SNR. We can directly see that the lowest $L_2$ error does not directly correspond to the highest SNR. Moreover, the visual evaluation suggests that the SNR describes the image quality more accurately.}
    \label{simimages}
\end{figure}

\begin{figure}
    \includegraphics[width = 1\textwidth]{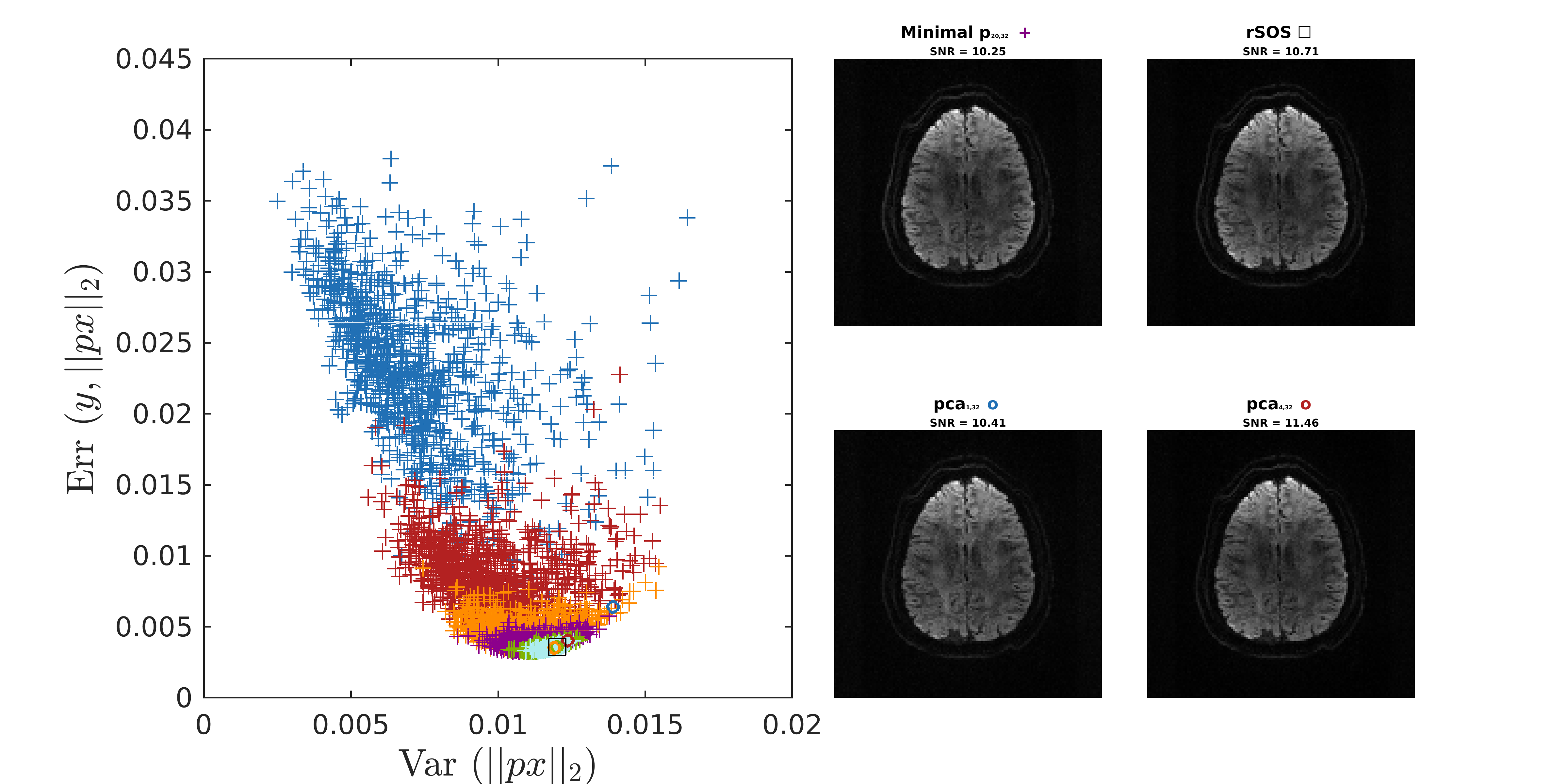}
    \caption{ Left - Scatter plot for the in-vivo data set showing the reconstruction error \eqref{err} and variance \eqref{var} obtained by combining the coils with rSOS and compressing with random projections and PCA in $\G_{k,32}$. Right - Final cross-sectional images corresponding to the compression methods in the scatter plot. The SNR value corresponds to the mean over all voxels in the reconstructed image volume, see \eqref{SNR}.}
    \label{invivo}
\end{figure}

Figure \ref{simplots} contains the scatter plots for $4$ different noise levels in the simulated data set and Figure \ref{simimages} shows corresponding image cross-sections. Figure \ref{invivo} shows the scatter plot and image cross-sections of the in-vivo data set.

\section{Discussion}\label{discussion}
In Figure \ref{simplots} and \ref{invivo} we illustrate the scatter plots of the described linear compression methods regarding the two data sets. To simplify the visualization we display the spaces $\G_{k,32}$ only for $k \in \{1,4,12,20,28,31\}$. In all plots we can see that the smaller the dimension $k$, the more the projections are spread regarding $\Err(y,\norm{px}_2)$ and $\Var\big(\norm{px}_2\big)$; the reconstructions including random projections from $G_{1,32}$ are widely distributed, whereas using the projections in $G_{31,32}$ are clustered closely around the rSOS. The smaller the $k$, the more original information can be randomly dismissed: the preservation and loss of the original information varies less for image space data compressed by a random projection $p \in G_{31,32}$ in comparison to compression by $p \in G_{1,32}$. 

For lower noise levels in the simulated data, Figure \ref{simplots} (a)-(b), we can directly see that the correlation between the $L_2$ reconstruction error and the variance within the final combined images is very strong. Since variance can be interpreted as contrast in images with high SNR, it shows that high contrast directly relates here to a good reconstruction in the $L_2$ manner. However, the higher the noise level, the more noise is also described in the measured variance and therefore it cannot directly be interpreted as contrast any more. In the simulated data set, Figure \ref{simplots} (c)-(d), this can be seen in the change of correlation behavior, where higher variance does not relate for all dimensions $k$ to a lower reconstruction error. In Figure \ref{simplots} (d) the projections from the spaces $G_{k,32}$  with $k = \{12,20,28,31\}$ do not show any connection between the variance and $L_2$ error. This might happen because the high noise level in the original data influences the measure of variance strongly, merging the underlying meaning of contrast. Interestingly, for $k=1$ there is still some negative correlation, indicating that the randomization leads to several poor reconstructions where noise is secondary in comparison to issues with contrast. The scatter plot corresponding to the in-vivo data (Figure \ref{invivo}) shows similar behavior. Because of this varying behavior, the variance itself does not act as a useful measure of reconstruction accuracy in this experimental setup.      

We can see that in all experiments rSOS itself, as well as including PCA compression, yield low $L_2$ reconstruction errors, but are always outperformed by some random projections. Nevertheless, regarding SNR and visualization, these reconstructed images with random compression cannot compete with PCA compression or rSOS. That indicates some contradicting behavior, when measuring the reconstruction performance with the $L_2$ error versus the SNR. Indeed, when using PCA for the simulated data set, the $L_2$ error is the lowest in $G_{31,32}$, whereas highest SNR is always achieved by PCA in $G_{4,32}$, which contradictory yields the worst $L_2$ error. Also in the in-vivo data set the highest SNR was achieved by using PCA in $G_{4,32}$, which again does not correspond to the lowest $L_2$ error. Following the visual results it seems that the SNR describes here the visual performance better than the $L_2$ error. The compression with PCA in $G_{4,32}$ yields the highest SNR in all experiments and therefore outperforms rSOS without compression consistently. Compression with the standard PCA in $G_{1,32}$ yields predominantly better results than rSOS, but yields an insufficient visual result on the in-vivo data. Just using the first eigendirection yields a strong compression that looses too much important information here.

\section{Summary}
Based on random orthogonal projections we have shown a numerical investigation on reconstruction accuracy regarding rSOS coil combination with linear compression. Two different MR data sets were used for our experiments; a simulated T1 weighted data set with varying amount of noise and an in-vivo diffusion weighted data set. We used diverse measures of performance to evaluate the image quality of the reconstructions. For the lower noise levels in the simulated data, the $L_2$ reconstruction error yields strong correlation with the variance, but the behavior changes for higher noise levels and the in-vivo data. Moreover, measuring $L_2$ error and SNR acts contradictory in terms of optimality and in these cases we observe that the visual evaluation corresponds more to the SNR. The highest SNR values were achieved by incorporating PCA as compression before using rSOS, outperforming rSOS with no compression in all experiments. This clearly suggests to use PCA on image space data before computing the final coil combination with rSOS, yielding a beneficial denoising effect with higher SNR. 

Future work shall include related experiments in the k-space before PI reconstruction, where linear compression is highly beneficial to save subsequent computational costs.

\section{Acknowledgment}
The work was partly funded by the Austrian Marshall Plan Foundation and Vienna Science and Technology Fund (WWTF) through project VRG12-009. 

\bibliographystyle{abbrv} 
{\footnotesize \bibliography{mri.bib}}

\end{document}